\documentclass[twocolumn,
showpacs,preprintnumbers,
showkeys,
amsmath,amssymb,
prb,reprint
]{revtex4-1}

\usepackage{graphicx}
\usepackage{dcolumn}
\usepackage{bm}
\usepackage{hyperref}
\usepackage{textcomp}

\begin{document}


\title{Spin transport in graphene nanostructures}

\author{M. H. D. Guimar\~aes}
 \email{m.h.diniz.guimaraes@rug.nl, guimaraes@cornell.edu}
\author{J. J. van den Berg}
\author{I. J. Vera-Marun}
\author{P. J. Zomer}
\author{B. J. van Wees}
\affiliation{Physics of Nanodevices, Zernike Institute for Advanced Materials, University of Groningen, The Netherlands}

\date{\today}

\begin{abstract}
Graphene is an interesting material for spintronics, showing long spin relaxation lengths even at room temperature.
For future spintronic devices it is important to understand the behavior of the spins and the limitations for spin transport in structures where the dimensions are smaller than the spin relaxation length.
However, the study of spin injection and transport in graphene nanostructures is highly unexplored.
Here we study the spin injection and relaxation in nanostructured graphene with dimensions smaller than the spin relaxation length.
For graphene nanoislands, where the edge length to area ratio is much higher than for standard devices, we show that enhanced spin-flip processes at the edges do not seem to play a major role in the spin relaxation.
On the other hand, contact induced spin relaxation has a much more dramatic effect for these low dimensional structures.
By studying the nonlocal spin transport through a graphene quantum dot we observe that the obtained values for spin relaxation are dominated by the connecting graphene islands and not by the quantum dot itself.
Using a simple model we argue that future nonlocal Hanle precession measurements can obtain a more significant value for the spin relaxation time for the quantum dot by using high spin polarization contacts in combination with low tunneling rates.

\end{abstract}

\maketitle

\section{Introduction}
Experiments and applications in the field of spin electronics, known as spintronics \cite{TregerScience2001}, often require that the spins keep their information for a long time, travel a long distance or that the devices show a large spin dependent signal.
Graphene has attracted a lot of attention for spintronics because of the long spin relaxation time ($\tau_{s}$) \cite{WeesNature2007,KawakamiPhys.Rev.Lett.2011,OezyilmazPhys.Rev.Lett.2011} which, in combination with its long mean-free path ($l_{mfp}$), leads to the longest spin relaxation length at room temperature ($\lambda_{s}$) \cite{WeesPhys.Rev.Lett.2014,arxiv-aachen}.
The spin accumulation in a device can be defined as \cite{WeesPhys.Rev.B2009} $\mu_{s} = (\mu_{\uparrow} - \mu_{\downarrow})/2$, where $\mu_{\uparrow (\downarrow)}$ is the chemical potential for spin up (down).
The relatively high sheet resistance ($R_{sq}$) of graphene flakes when compared to normal metals, and its robustness to high current densities \cite{BachtoldPhys.Rev.Lett.2009}, makes graphene an efficient system for the creation of large spin accumulation \cite{WeesPhys.Rev.B2011,ValenzuelaAppliedPhysicsLetters2013} which results in large spin dependent signals.

A successful route for increasing the spin accumulation in a device without the need of large current densities is by miniaturizing the devices to scales where their width $W$ and length $L$ are smaller than $\lambda_{s}$ \cite{WeesPhys.Rev.Lett.2003,WeesPhys.Rev.B2005,FertPhys.Rev.B2010,FertPhys.Rev.B2012}.
At these scales the spin accumulation is confined in a small area and the spins do not diffuse away as in the case of standard devices \cite{WeesPhys.Rev.B2014}.
A twofold increase in the local spin signal was demonstrated in metallic devices where $L > \lambda_{s}$ \cite{FertPhys.Rev.B2012}.
Since graphene has $\lambda_{s} \approx$ 2 \textmu m \cite{WeesPhys.Rev.B2009}, the fabrication of devices with dimensions smaller than $\lambda_{s}$ is much easier than for metals, where $\lambda_{s}$ is usually smaller than 1 $\mu$m \cite{WeesNature2002}.
Therefore, the increase in spin signal is expected to be even larger for graphene based nanodevices.
Moreover, it was already demonstrated that the spin signal can be further increased due to quantum interference effects in graphene devices where the phase coherence length is smaller than the device dimensions \cite{Guimaraes2014}, in a similar way to carbon nanotube devices \cite{Feuillet-Palma2010,Man2006}.

However, it is possible that the edges of the graphene flake have a limiting influence for the spin relaxation \cite{WeesPhys.Rev.B2009,Dugaev2014}.
It is known that edge states in graphene can be spin polarized \cite{LouiePhys.Rev.Lett.2006,CrommieNatPhys2011,SchmidtPhys.Rev.Lett.2014} which can enhance spin-flip processes at the edges \cite{Dugaev2014}.
By lowering the dimensions of the graphene flake to below $\lambda_{s}$, the edge length to area ratio increases and the spins probe the edges of the structure more often than for a regular size device.
Therefore, the study of graphene spin valves with small dimensions gives insights about the role of the edges on the spin relaxation.

When the device dimensions are even smaller, in the order of $l_{mfp}$, quantum confinement of the electrons can be obtained \cite{Stampfer2008,EnsslinMaterialsToday2010}.
For electrons confined in three dimensions we have the solid state analogous to an atom, a quantum dot (QD), which shows discrete energy levels \cite{TaruchReportsonProgressinPhysics2001}.
Spins in a QD are heavily used for quantum information processing and quantum computation using spin qubits \cite{VandersypenRev.Mod.Phys.2007}.
Graphene quantum dots are predicted to have spin relaxation times two orders of magnitude longer than for pristine graphene flakes \cite{BurkardNatPhys2007}, which makes graphene very appealing for quantum computation.
However, the study of spin relaxation in quantum dots is not trivial, usually demanding fast and precise voltage sources and/or the fabrication of two coupled quantum dots \cite{VandersypenRev.Mod.Phys.2007}.
The possibility of studying the spin transport properties of QDs using nonlocal techniques is therefore an appealing alternative.

In order to obtain effective electric spin injection into graphene we have to overcome the issue known as the conductivity mismatch problem \cite{WeesPhys.Rev.B2000,RashbaPhys.Rev.B2000}.
This issue arises because the spin resistance of graphene $R_{\lambda}$ is generally much higher than the spin resistance of the ferromagnetic metals used for spin injection $R_{\lambda_{F}}$.
If the ratio $R_{FM} / R_{\lambda}$ is much smaller than 1, the spins tend to return to the ferromagnetic contacts and relax there.
For a graphene flake much longer than the spin relaxation length ($L \gg \lambda_{s}$), the spin resistances for the graphene flake and a ferromagnet contacting it can be defined as: $R_{\lambda} = R_{sq} \lambda_{s} / W$ and $R_{\lambda_{F}} = \rho_{FM} \lambda_{FM} / A$, where $W$ is the graphene channel width, and $\rho_{FM}$, $\lambda_{FM}$ and $A$ are, respectively the resistivity, spin relaxation length and cross sectional area of the ferromagnet.
Generally the spin relaxation length in graphene is much longer than for the ferromagnet, and graphene's resistivity is usually much higher than the resistivity of ferromagnetic metals, leading to a ratio $R_{FM} / R_{\lambda} \ll 1$, and resulting in a poor spin injection.
This problem can be circumvented by the use of highly resistive barriers, where the term $R_{FM}$ is then substituted by the contact interface resistance $R_{c}$.
The problem of conductivity mismatch in the context of graphene spintronics has received much attention in the past years \cite{WeesPhys.Rev.B2009,WeesPhys.Rev.B2012,KawakamiPhys.Rev.Lett.2010,Dlubak2012}.
Estimations of the influence of the contacts in the measured spin relaxation in graphene show that most of the measurements performed in a nonlocal 4-probe geometry are not limited by contact induced spin relaxation \cite{WeesPhys.Rev.B2012,KawakamiPhys.Rev.Lett.2010}.
On the other hand, local 2-probe measurements where the contact resistance is orders of magnitude higher than for conventional devices estimate a much higher spin relaxation time in epitaxial graphene on SiC\cite{Dlubak2012}.
However, the influence of localized states in the SiC substrate \cite{Maassen2013} and the fact that spin signals were only reported at 4.2 K hinders the comparison with other experiments.

Here we report spin injection and transport in graphene nanostructures with dimensions smaller than $\lambda_{s}$.
By Hanle precession measurements we obtain the spin relaxation time in these devices $\tau_{s} \approx 30$ ps.
Using a model that takes into account the size of the devices, we show that for graphene nanoislands where $L,W < \lambda_{s}$ the nonlocal spin signal can be increased by a factor of $\approx 100$ when compared to standard devices, where $L \gg \lambda_{s}$.
Our simulations show that contact induced spin relaxation effects have a larger influence on measurements in confined devices than in devices where the length is longer than $\lambda_{s}$.
Comparing our simulations with the experiments we find that the experimentally obtained values for $\tau_{s}$ are limited by contact induced spin relaxation.
An estimation of the intrinsic values for the spin relaxation times in our graphene nanoislands results in $\tau_{s} \approx$ 200 ps, indicating that the enhancement of spin-flip processes at the edges is not the main mechanism for spin relaxation in graphene on SiO$_{2}$.

Furthermore, we study the spin transport through an open graphene quantum dot, which consists of a graphene quantum dot connected by two graphene nanoislands from each side.
We show that the measured spin relaxation is dominated by the graphene areas that connect the quantum dot in the case where the tunneling rate to the dot is high.
The transition between the spin relaxation happening mostly in the quantum dot to spin relaxation happening mostly in the outside areas is explored as a function of the tunneling rate to the dot using a simple model.

\section{Methods}
Our samples are prepared using mechanically exfoliated graphene on 500 nm SiO$_{2}$/Si substrates.
Single layer graphene flakes were selected using optical contrast and confirmed by atomic force microscopy (AFM).
In order to increase the precision for the next fabrication steps we use electron beam lithography (EBL) to define alignment markers close to the selected flake.
In the same EBL step we pattern bonding pads and large contact wires and finally evaporate Ti/Au (5/35 nm) by the use of an electron beam induced evaporator.
To pattern our graphene flakes into the structures used here we perform another EBL step to define an etching mask using high molecular weight (950K) polymethyl methacrylate, and use a pure oxygen plasma to etch the graphene flakes.
After the etching procedure the structures are once again analyzed by the use of AFM to select the structures with cleaner surfaces and ensure a homogeneous contact interface for the electrodes.
The electrodes are patterned by another EBL step followed by metal evaporation.
To avoid the conductivity mismatch problem we first fabricate a TiO$_{2}$ layer by evaporating 0.4 nm of Ti followed by oxidation in a pure oxygen atmosphere at pressures above 10$^{-1}$ Torr.
This step is repeated once more and, after the evaporation chamber is pumped to high vacuum, 35 nm of Co are evaporated.
All the electrical measurements here are performed using standard low-frequency ($f < 20$ Hz) lock-in techniques with bias current between 10 nA and 2 $\mu$A.

\section{Graphene nanoislands}
\subsection{Experiment}
We start discussing the experimental results for graphene nanoislands in which $L,W < \lambda_{s}$.
We studied a total of 3 devices of this type.
Two of the devices with dimensions $1 \times 0.5$ $\mu$m$^{2}$ and one with $1 \times 0.25$ $\mu$m$^{2}$.
All contact resistances for these 3 devices are between $R_{c} =$ 10 and 36 k$\Omega$.
While for some devices the contact resistances are in the order of tens of k$\Omega$, for others they can reach above M$\Omega$.
We attribute this to undesired contamination resulting from the etching procedure, as observed before \cite{WeesPhys.Rev.B2009}.
Unless specified otherwise, all the results reported here were obtained at room temperature.

It is important to note that the edges of etched flakes are usually irregular at the atomic scale, not following a specific crystalographic orientation, due to the rough etching procedure.
However, even though the rough edges do not have all the characteristics of a crystallographic edge, they can show localization and enhanced scattering, which can enhance spin-flip processes \cite{Dugaev2014}.

\begin{figure}[h]
	\centering
		\includegraphics[width=0.50\textwidth]{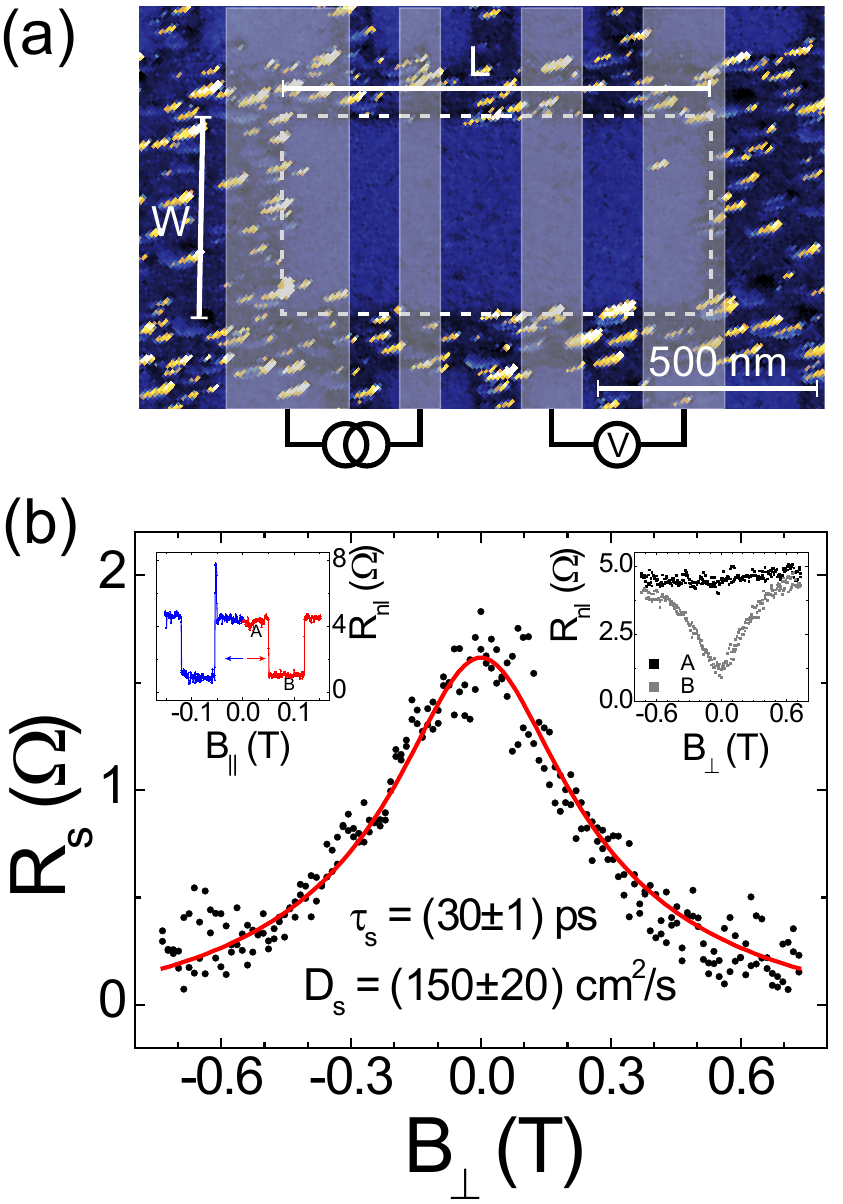}
	\caption{(Color online) (a) Phase contrast atomic force micrograph of a typical graphene nano-island device. The graphene is outlined by the dashed line and the contacts are represented by the lighter semi-transparent blocks. The measurement scheme for nonlocal spin transport is shown below. (b) Nonlocal spin signal as a function of a perpendicular magnetic field. 
The line in red is a fit using Eq. \ref{eq:oned} to extract the values for $D_{s}$ and $\tau_{s}$. \textit{Left inset:} Nonlocal spin signal as a function of a parallel magnetic field. The arrows indicate the direction of the magnetic field sweep. \textit{Right inset:} Hanle precession for $R_{A (B)}$ in black (grey).}
	\label{fig:island}
\end{figure}

Fig. \ref{fig:island} shows a phase contrast AFM image of one of the devices before contact deposition with the contact pattern outlined by lighter semi-transparent blocks.
In order to perform the spin dependent measurements with minimum contribution of the charge dependent signal we use the nonlocal technique \cite{WeesPhys.Rev.B2009}, Fig. \ref{fig:island}(a).
When a current $I$ is driven between two ferromagnetic electrodes, a spin accumulation is generated underneath the injection electrodes which diffuses away from the injection point.
The voltage $V_{nl}$ is probed outside the charge current path, which minimizes the charge contribution to the signal and, since the voltage probes are also ferromagnetic, the chemical potential for a specific spin species is preferentially detected.

For devices like the ones shown here, the spins can probe the whole flake before relaxing which causes the spin accumulation in the device to be, in principle, homogeneous \cite{WeesPhys.Rev.Lett.2003,WeesPhys.Rev.B2014}.
The presence of a spin accumulation can be tested by a nonlocal spin valve measurement where a large negative parallel magnetic field $B_{||}$ is applied to the device followed by a sweep in $B_{||}$ while recording the nonlocal resistance, $R_{nl}=V_{nl}/I$.
Since the electrodes have different widths, their coercive fields vary, which causes their magnetization to switch direction at different values of magnetic field.
The switches in magnetization of the electrodes can be seen as abrupt steps in the nonlocal resistance [left inset of Fig. \ref{fig:island}(b)].

In order to extract the spin relaxation time in our devices we perform Hanle precession measurements where a perpendicular magnetic field $B_{\bot}$ is applied to the device causing the injected spins to precess around the field.
This results in a decrease of the spin signal due to incoherent spin precession and spin relaxation.
The resultant signal can then be fitted by the solution to the Bloch equations for spin diffusion:

\begin{equation}
D_{s} \frac{d^{2} \vec{\mu}_{s}}{dx^{2}} - \frac{\vec{\mu}_{s}}{\tau_{s}} + \omega \times \vec{\mu}_{s} = 0,
 \label{eq:bloch}
\end{equation}

\noindent where $D_{s}$ is the spin diffusion constant, $\omega=g \mu_{B} \vec{B}/ \hbar$ with $g$ the Land\'e g-factor, $\mu_{B}$ the Bohr magneton and $\vec{B}=B_{\bot}\hat{z}$ the magnetic field.
The spin relaxation length can be calculated using the relation $\lambda_{s} = \sqrt{D_{s} \tau_{s}}$.

Considering non-invasive contacts and a one-dimensional (1D) infinite channel, the solution for Eq. \ref{eq:bloch} for the spin signal as a function of $B$ for one spin injector and one spin detector is \cite{Jedema2002}:

\begin{equation}
R_{s} = \frac{P^{2} R_{sq} D_{s}}{W} \int_{0}^{\infty}{\mathcal{P}(t) cos(\omega t) exp(-t/ \tau_{s}) dt},
\label{eq:oned}
\end{equation}

\noindent where $\mathcal{P}(t) = (4 \pi D_{s} t)^{-1/2} exp(-\ell^{2} / 4 D_{s} t)$, $\ell$ is the contact spacing, and we assumed that the spin polarization $P$ of the spin injector and detector electrodes are equal.

Fitting the experimental data with Eq. \ref{eq:oned} allows us to obtain $\tau_{s}$ and $D_{s}$ independently.
For the zero-dimensional (0D) case, considering uniform spin accumulation, the spin signal has a Lorentzian form \cite{WeesPhys.Rev.Lett.2003}:

\begin{equation}
R_{s} = \frac{P^{2}}{e^{2} \nu_{DOS} A} \left( \frac{\tau_{s}}{1 + (\omega \tau_{s})^{2}} \right),
\label{eq:zerod}
\end{equation}

\noindent where $e$ is the electron charge, $\nu_{DOS}$ is the density of states, and $A$ the area of the island.

In order to avoid contribution from background signals to our analysis \cite{WeesPhys.Rev.B2012}, we calculate the spin dependent signal as: $R_{s} = (R_{A} - R_{B}) / 2$, where $R_{A(B)}$ are the values obtained for Hanle precession measurements at the magnetization configuration of the electrodes specified as $A$ ($B$) \footnote{Note that the Hanle precession measurement for the ‘all parallel’ configuration ($A$) is independent of magnetic field. This is due to the fact that the injection electrodes are close together and, while one injects spin up, the other extracts the same spin species at a very close location, leading to a low total spin accumulation. Furthermore, the two detection electrodes probe the same spin species almost at the same location, leading to a lack of sensitivity to the spin accumulation.}. For the case of the measurement shown in Fig. \ref{fig:island}(b), we have that $A$ is the configuration where all electrodes are aligned parallel to each other and in configuration $B$ the inner injector is aligned antiparallel to the other three electrodes.

By fitting the results for all three studied graphene nanoisland devices, we obtain that $\tau_{s}$ falls in the range 10 - 30 ps with little variation between fitting the data with Eq. \ref{eq:oned} or with Eq. \ref{eq:zerod}. The values of $\tau_{s}$ obtained by the use of Eq. \ref{eq:zerod} are systematically lower by a factor $\approx$ 2. This difference is in agreement with previously reported results \cite{WeesPhys.Rev.B2014} showing that Eq. \ref{eq:oned}, which disregards reflection of the spin accumulation at the edges, results in an overestimation of $\tau_{s}$ by a factor $\approx$ 2 for very small systems. $D_{s}$ obtained by the use of Eq. \ref{eq:oned} in our devices varies from 0.01 to 0.001 m$^{2}$/s. The fact that we obtain values for $D_{s}$ which are comparable to the standard values obtained on devices where $L \gg \lambda_{s}$, $D_{s} \approx 0.02$ m$^{2}$/s, is an indication that the spin accumulation in our devices is not truly 0D \cite{WeesPhys.Rev.B2014}. As it will be clarified below by the use of a model that includes the effects of the electrodes on the spin relaxation, since the contact resistance for our devices is small, the contacts reduce the spin accumulation underneath them due to contact induced spin relaxation. This results in a gradient on $\mu_{s}$ throughout the nanoisland. From now on, when the values for $\tau_{s}$ and $D_{s}$ are discussed we will use the values obtained using Eq. \ref{eq:oned}, unless specified otherwise.

The obtained $\tau_{s}$ for our devices is one order of magnitude lower than for regular SiO$_{2}$ based graphene devices where $\tau_{s} \approx$ 200 ps \cite{WeesPhys.Rev.B2009} and about two orders of magnitude below the best graphene spintronics devices where $\tau_{s} \approx 1$ ns \cite{KawakamiPhys.Rev.Lett.2011,WeesPhys.Rev.Lett.2014,arxiv-aachen}.
Although this difference could be explained by enhanced spin-flip processes at the edges, we will show below that the most probable cause is a demeaning influence of the contacts on the spin transport.
Since the spins are confined to the graphene island and the contacts cover more than half the area of the device, the influence of the contacts in the spin relaxation is expected to be much larger than for regular devices where the area covered by the contacts is less than 10$\%$ the total device area.
It is important to note that the spin signal obtained in our graphene nanoislands is considerably smaller than the values predicted in the introduction. As we will demonstrate in the next section, this also results from the low contact resistance in our devices.
To understand and explain the physics behind our experiments we model our system as explained below.

\subsection{Simulations}
\begin{figure}[h]
	\centering
		\includegraphics[width=0.50\textwidth]{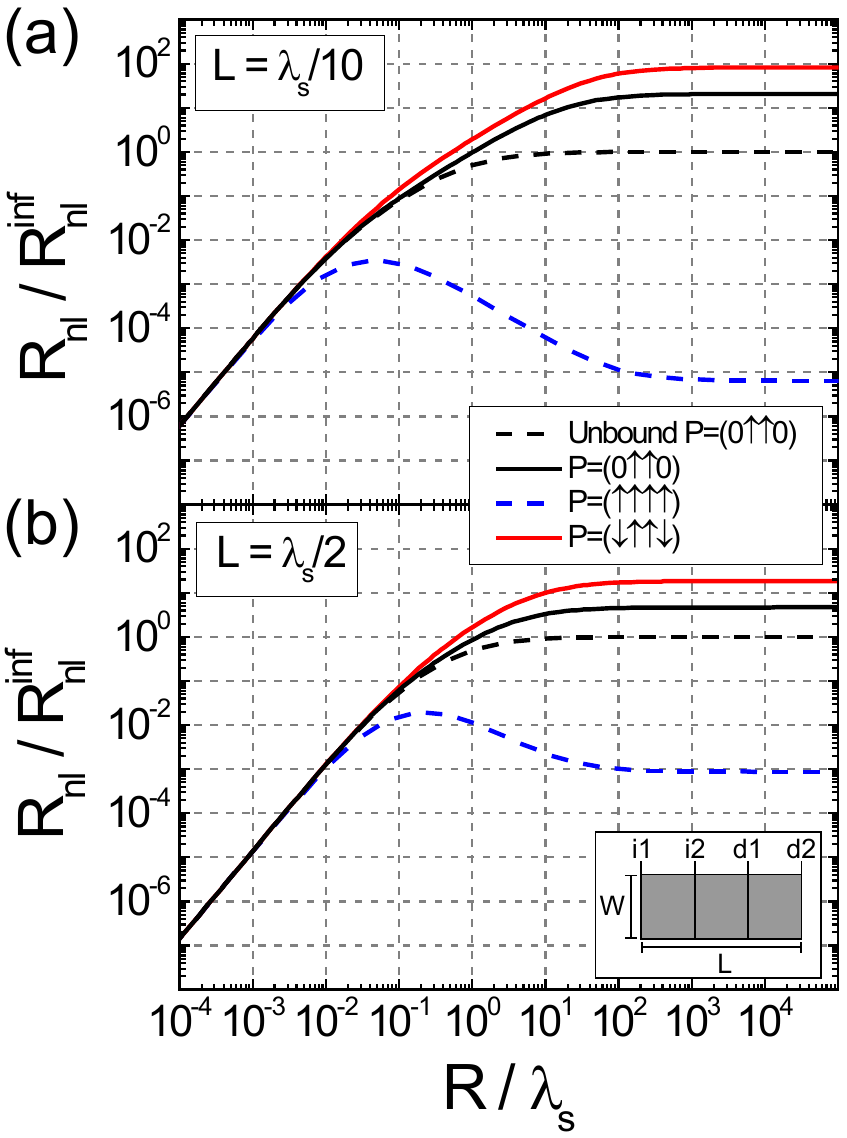}
	\caption{(Color online) Nonlocal resistance normalized by the maximum value obtained in an infinite (unbound) system as a function of the ratio $R / \lambda_{s}$ for systems of length (a) $L=\lambda_{s} / 10$ and (b) $L=\lambda_{s} / 2$. The values considering one spin injector and one spin detector aligned in parallel are shown by the solid black lines. The case of two spin injectors and two spin detectors in parallel and antiparallel alignment are represented by the dashed blue and solid red lines respectively. For comparison, the case of an unbound system with one spin injector and one detector is represented by the dashed black line. The schematics of the system showing the dimensions and the four contacts is shown in the inset of (b).}
	\label{fig:islandsim1}
\end{figure}

We use the well established diffusive model to describe the motion and relaxation of spins in our system including contact induced spin relaxation effects \cite{WeesPhys.Rev.B2009,WeesPhys.Rev.B2012}.
The system has a width $W$ and length $L$ with 4 contacts spaced by $\ell=L/3$, two injection contacts, $i1$ and $i2$ and 2 detection contacts, $d1$ and $d2$ [inset of Fig. \ref{fig:islandsim1}(b)].
For the simulations we fix the values for the spin relaxation time $\tau_{s}$ = 200 ps and the spin diffusion constant $D_{s} = 0.02$ m$^{2}$/s for the system.
This results in a spin relaxation length of $\lambda_{s}$ = 2 $\mu$m.
We solve Eq. \ref{eq:bloch} with the boundary conditions that the spin accumulation is continuous everywhere and that the spin current is continuous inside the system except at the injection points.
Here we include a source term $P_{i} I / W$, where $P_{i}$ is the spin injection efficiency.
The spin current is set to be zero at the boundaries $x=0$ and $x=L$.
The contact induced spin relaxation is included in the same way as in Ref. \cite{WeesPhys.Rev.B2009,WeesPhys.Rev.B2012}.
It is important to point out that this model is valid for contacts with low spin polarization, as in the case of our experiments, where $P \approx 0.1$.

In order to quantify the influence of the contact resistance in the spin transport we compare the parameter $R = R_{c} W / R_{sq}$ with the spin relaxation length \cite{WeesPhys.Rev.B2009,WeesPhys.Rev.B2012}.
For $R / \lambda_{s} \ll 1$ the spins tend to go back to the contacts and relax due to the shorter spin lifetime in the ferromagnetic metal.
In the case $R / \lambda_{s} \gg 1$ the contacts are non-invasive and do not disturb the spin accumulation underneath.
We start by comparing the nonlocal resistance as a function of the ratio $R / \lambda_{s}$ for a system with total length $L = \lambda_{s} / 10$ with an unbound system (infinite length), Fig. \ref{fig:islandsim1}(a).

The spin polarization of the contacts is taken to be: +0.1 (represented by $\uparrow$), -0.1 (represented by $\downarrow$) or 0 (no spin polarization).
To summarize the polarization of the 4 contacts involved we use $P=(P_{i1},P_{i2},P_{d1},P_{d2})$.
We will consider three different cases: one spin injector and one spin detector [$P=(0,\uparrow,\downarrow,0)$] in an unbound system, one spin injector and one spin detector in a finite device of length $L$, and two spin injectors and two spin detectors [$P=(\uparrow,\uparrow,\uparrow,\uparrow)$ and $P=(\downarrow,\uparrow,\uparrow,\downarrow)$] in a finite device with length $L$.
The first case is shown as a dashed black line in Fig. \ref{fig:islandsim1}.
We see that our curve agrees with previously reported results \cite{WeesPhys.Rev.B2009,WeesPhys.Rev.B2012}, with the nonlocal signal reaching approximately 90$\%$ of the maximum signal at $R / \lambda_{s} \approx$ 10.
For the second case, the bound system with one spin injector and one detector (solid black line), we observe that the maximum nonlocal signal is about one order of magnitude higher than for the unbound system.
This increase in the spin accumulation is due to reflection of the spins at the boundaries and also agrees with previously reported results \cite{WeesPhys.Rev.B2014}.
The point where the nonlocal signal is 90 $\%$ of the maximum signal is at $R / \lambda_{s} \approx$ 200, more than one order of magnitude higher than the one for the infinite system.
This can be understood by the fact that since the backflow of spins into the ferromagnet (across the tunnel barrier) is driven by the spin accumulation underneath, an increase in $\mu_{s}$ due to the confinement results in an increase in the backflow of spins.
This picture can be alternatively viewed as an increase of the effective spin resistance in graphene due to confinement.

When two spin injectors and two detectors are considered, we observe that when the contacts are all in a parallel alignment [$P=(\uparrow,\uparrow,\uparrow,\uparrow)$, dashed blue line] the nonlocal signal is very small, orders of magnitude lower when compared with the previous case.
This is due to two factors.
First, the contacts are separated by less than $\lambda_{s}$.
While one of the injector contacts \textit{injects} spin up, the other \textit{extracts} spin up at the same rate given their equal polarization.
Furthermore, both detection contacts probe approximately the same value of chemical potential.
Second, the spin accumulation is approximately constant throughout the system due the 0D behavior of $\mu_{s}$ \cite{WeesPhys.Rev.Lett.2003}, which enhances the effects described above for the injection and detection circuits.
It is important to note that these effects are reduced when the length of the device and spacing between contacts is increased, as can be seen when comparing Fig. \ref{fig:islandsim1} (a), which has $L$ = $\lambda_{s}$/10 and (b) with $L$ = $\lambda_{s}$/2.

The highest spin signal in the case of two spin injectors and two detectors is obtained when both injector and detector pairs have an anti-parallel alignment [$P=(\downarrow,\uparrow,\uparrow,\downarrow)$, solid red line].
In this case, while one injector injects spin up, the other extracts spin down, which increases the total spin accumulation in the device by a factor $\approx$ 2 when compared to the case of only one spin injector.
Moreover, one of the detectors is sensitive mostly to the chemical potential for spin up, while the other detector senses spin down, this gives another factor of $\approx$ 2, resulting in a total increase of $\approx$ 4 to the spin signal when comparing to the case of one spin injector and one spin detector.

Since the spin accumulation in the case of 4 contacts with non-zero spin polarization is higher than when considering just 2 contacts, the saturation of the signal ($90\%$ of the maximum signal) only occurs at $R / \lambda_{s} \approx$ 400, twice the value found for the case of one injector in a closed system and about 40 times higher than for an unbound system.
This indicates that a high quality resistive interface is very important when studying the spin transport in confined geometries.
This fact was shown by Laczkowski et al. \cite{FertPhys.Rev.B2012} with results obtained by a transfer matrix technique.

In order to compare the results of the simulations to our devices, we use the dimensions of our samples $L=1$ $\mu$m and $W=0.5$ $\mu$m, and assume a standard value for the spin relaxation in graphene $\lambda_{s} \approx$ 2 $\mu$m, resulting in $2L=\lambda_{s}$, Fig. \ref{fig:islandsim1}(b).
The effects of confinement in this case are less pronounced than for those shown in Fig. \ref{fig:islandsim1}(a).
The increase of the spin signal due to confinement, even when considering the contribution of all 4 contacts is only about a factor of 10.
This results in a smaller, but still noticeable influence of the contact resistance on the spin relaxation, showing saturation ($90\%$) of $R_{nl}$ for values above $R / \lambda_{s} \approx 100$ when considering 4 contacts with non-zero spin polarization.

\begin{figure}[h]
	\centering
		\includegraphics[width=0.50\textwidth]{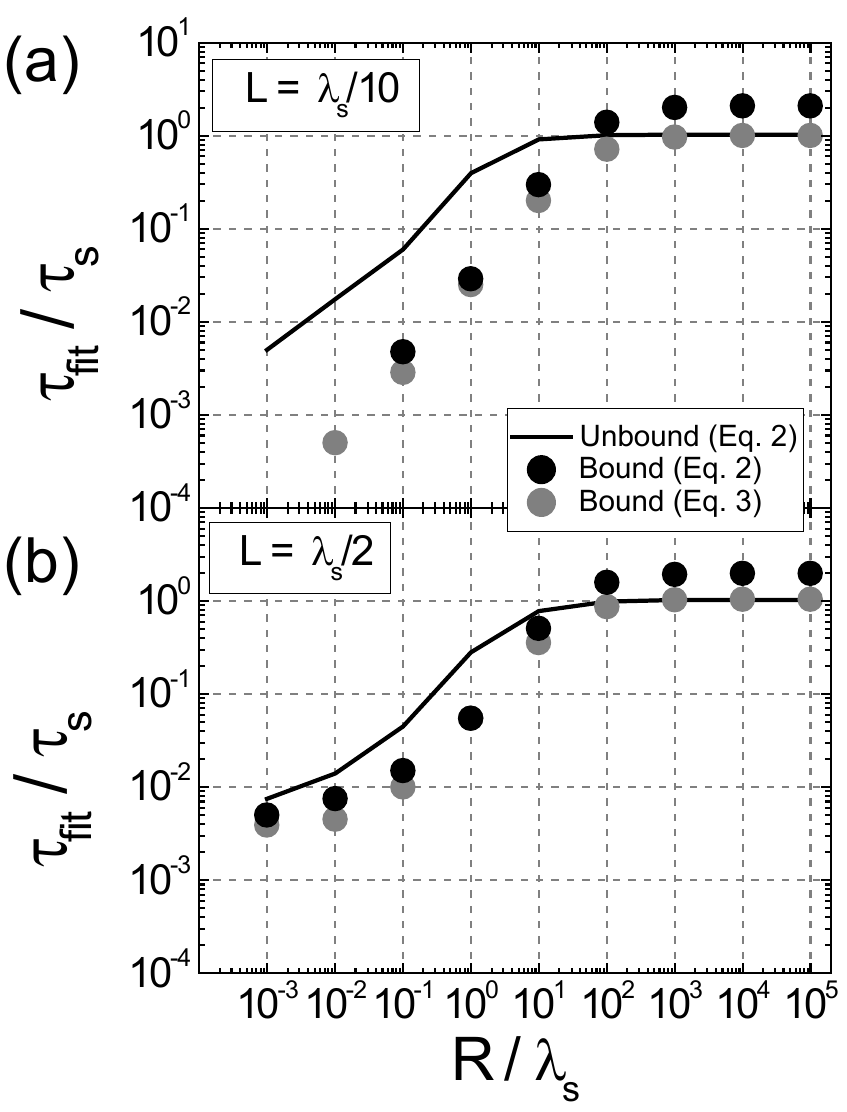}
	\caption{(Color online) The ratio $\tau_{fit} / \tau_{s}$ as a function of $R / \lambda_{s}$ for finite systems (bound systems) of length (a) $L=\lambda_{s} / 10$ and (b) $L=\lambda_{s} / 2$. The results obtained by fitting the simulated Hanle precession curves using Eq. \ref{eq:oned} are represented by the black circles and the results obtained by fitting using Eq. \ref{eq:zerod} are shown by the grey circles. The solid black line shows the results for the infinite (unbound) system using Eq. \ref{eq:oned}.}
	\label{fig:islandsim2}
\end{figure}

However, we are not only interested in explaining changes in magnitude of the signal but also in understanding how the Hanle precession measurements are affected by the presence of the contacts.
Therefore, we simulate Hanle precession curves using the same model described above for known values of $\tau_{s}$.
For comparison we also generate the data for the unbound system as studied by Maassen et al. \cite{WeesPhys.Rev.B2012}.
As in the experiments, the Hanle precession curves can be fitted by Eq. \ref{eq:oned} or by Eq. \ref{eq:zerod} to obtain a value for the spin relaxation time, $\tau_{fit}$.
In order to compare the values extracted by the fitting procedures and the values used in the simulation for the spin relaxation time, we take the ratio: $\tau_{fit} / \tau_{s}$.
The Hanle curves simulated for the unbound (infinite) system are fitted using only Eq. \ref{eq:oned} and the ones generated for the confined system are fitted with both models, Eqs. \ref{eq:oned} and \ref{eq:zerod}, for comparison (Fig. \ref{fig:islandsim2}).

All systems follow a similar trend of $\tau_{fit}$ as a function of the ratio $R / \lambda_{s}$: $\tau_{fit}$ increases with increasing $R / \lambda_{s}$ and after a certain value for this ratio $\tau_{fit}$ saturates.
We find that, as the experimental results shown in the previous section, the values for $\tau_{fit}$ obtained by Eq. \ref{eq:oned} is about a factor of 2 higher than the values obtained when we fit the Hanle precession curves with a Lorentzian (Eq. \ref{eq:zerod}).
By comparing Fig. \ref{fig:islandsim2} (a) and (b) we can see how confinement affects the extraction of the spin relaxation time as a function of the contact resistance.
For systems where $L \ll \lambda_{s}$ [Fig. \ref{fig:islandsim2}(a)], the value for the spin relaxation time extracted by fitting the Hanle precession curves only saturates around $R / \lambda_{s} \approx$ 100.
When $L = \lambda_{s} / 2$ [Fig. \ref{fig:islandsim2}(b)], the saturation of the extracted spin relaxation time happens around $R / \lambda_{s} \approx$ 10.
It is important to notice that in both cases, $L = \lambda_{s} / 10$ and $L = \lambda_{s} / 2$, the saturation of $\tau_{fit} / \tau_{s}$ occurs for values larger than those for the unbound case (black solid line).
As in the case of the nonlocal signal (Fig. \ref{fig:islandsim1}), this is due to the enhanced backflow of spins due to the higher $\mu_{s}$ for confined systems.

In order to give an estimation of the actual spin relaxation time in our experiments shown in the previous section, we use the experimental values $L$ = 1 $\mu$m, $W \approx$ 0.5 $\mu$m, and $R_{c}$ = 10 - 36 k$\Omega$.
With $R_{sq} \approx$ 1 k$\Omega$, we have that $R \approx$ 5 - 18 $\mu$m.
Using a standard value for the spin relaxation length of SiO$_{2}$ based graphene devices, $\lambda_{s} \approx$ 2 $\mu$m, we have that $R / \lambda_{s} \approx$ 2.5 - 9.
In Fig. \ref{fig:islandsim2}(b) we can see that such a range for $R / \lambda_{s}$ results in $\tau_{fit} \approx 0.1 \tau_{s}$.
This means that, for a standard value for graphene of $\tau_{s} \approx$ 200 ps, the spin relaxation time obtained from fitting the Hanle precession curves is $\tau_{fit} \approx 20$ ps, precisely in the range encountered in our experiments.
Therefore we can conclude that our experimental results are still dominated by contact induced spin relaxation and further improvement of the contact interface is necessary to unveil the full potential of confinement in the spin signal in graphene nanodevices.

As an extra confirmation of this last result we simulate our system using the experimental values and, instead of varying the ratio $R / \lambda_{s}$ we change the value for $\tau_{s}$.
We observe that $\tau_{fit}$ increases monotonically and saturates at $\tau_{fit} \approx$ 20 ps for values of $\tau_{s}$ above 100 ps.
Therefore we can conclude that the values for $\tau_{s}$ for the graphene nanoislands in our experiments is considerably larger than $\tau_{s}$ = 20 ps.
Although we cannot rule out completely the effect of edge scattering on the spin relaxation in our devices, we can state that if edge scattering does have an effect on the spin relaxation, it is not the dominant mechanism for SiO$_{2}$ based graphene devices since we do not observe any decrease on $\tau_{s}$ for our nanoislands with an increased edge length to area ratio.


\section{Graphene quantum dots}
\subsection{Experiment}
As mentioned in the introduction, the experimental determination of the spin relaxation time in QDs is a difficult task often requiring very complicated techniques.
An easier method to extract $\tau_{s}$ in QDs would be by the use of Hanle precession techniques.
However, QDs are often measured in a 2 probe configuration with non-spin polarized contacts, which makes it especially difficult to detect spin precession signals.

\begin{figure}[h]
	\centering
		\includegraphics[width=0.50\textwidth]{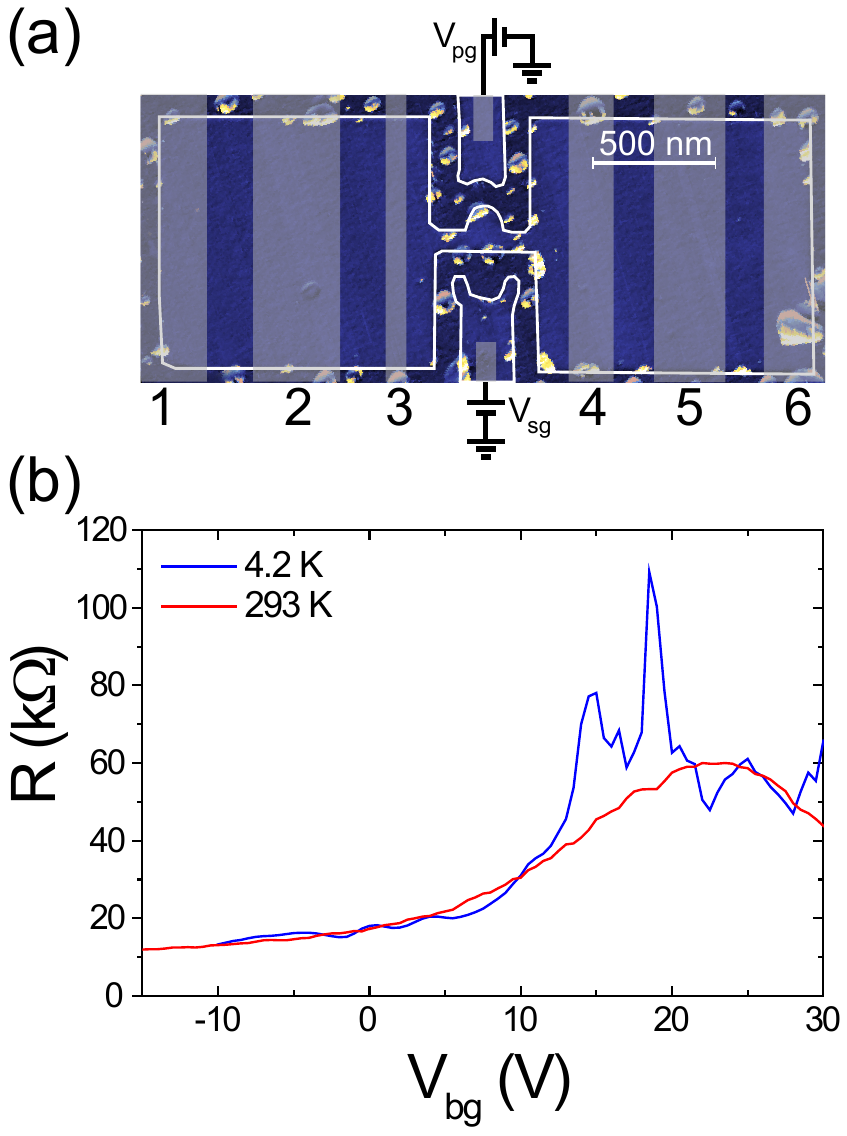}
	\caption{(Color online) (a) Phase contrast atomic force micrograph of a graphene quantum dot device. The graphene is outlined by the solid line and the contacts are represented by the lighter semi-transparent blocks. The electrodes are numbered 1 - 6. (b) Four-terminal resistance as a function of $V_{bg}$ with $V_{pg}=V_{sg}=$ 0 V for 293 K (red) and 4.2 K (blue). For this measurement a current was driven between electrodes 1 and 5, and the voltage detected between electrodes 3 and 4.}
	\label{fig:qd}
\end{figure}

Here we investigate the spin relaxation time in graphene QDs using a nonlocal technique that separates the charge and spin contribution.
We study the spin transport through the device in the presence of a perpendicular magnetic field in order to obtain information on the spin dynamics, e.g. the spin relaxation time.
A phase contrast AFM image of one of our devices is shown in Fig. \ref{fig:qd}(a).
The graphene structure consists in two graphene islands with dimensions 1x1 $\mu$m$^{2}$ connected by a QD defined as two narrow ($\approx$ 80x100 nm$^{2}$) constrictions with a broader region ($\approx$ 150 nm) in the center.
Two additional graphene structures disconnected from the rest serve as side-gate (sg) and plunger-gate (pg) electrodes to locally tune the chemical potential in the constrictions and QD, respectively.

In order to characterize the charge properties of the device we perform a four terminal measurement where the current is driven between two outer electrodes (1 and 5) and the voltage detected between the electrodes close to the QD (3 and 4).
Fig. \ref{fig:qd}(b) shows the four-terminal resistance of the quantum dot at room temperature and 4.2 K as a function of $V_{bg}$ for the plunger-gate and side-gate voltages set to $V_{pg}=V_{sg}=0$ V.
Although we could not reach the Coulomb blockade regime, the low temperature resistance curve shows peaks and dips which are indications of confinement in the structure \cite{TaruchReportsonProgressinPhysics2001}.

The spin transport experiments were carried out using the nonlocal geometry where the current is driven between electrodes 3 and 1 and the voltage is detected outside the current path, between electrodes 4 and 5 [Fig. \ref{fig:qd}(a)].
The contact resistances for this sample are in the range 100 - 700 k$\Omega$, considerably larger than the previous ones.
Using a standard value for the spin relaxation length on SiO$_{2}$ based devices of $\lambda_{s}$ = 2 $\mu$m and the values for the square resistance of the graphene islands connecting the dot $R_{sq} \approx$ 2 k$\Omega$ \footnote{Here we use the values for $R_{sq}$ for the regions underneath of the injection points, which in this case is given by $R_{sq}$ of the graphene islands that connect the QD.}, we have $R / \lambda_{s} \approx$ 100 - 700.
As discussed in the previous section, for these values of $R / \lambda_{s}$ contact induced spin relaxation effects do not play a major role in the measurements.
In this case, $R_{nl}$ and $\tau_{fit}$ are above 90$\%$ of the intrinsic values even for confined geometries.
Since the contacts do not induce extra relaxation, we expect the spin accumulation in the graphene islands to be constant.

\begin{figure}[h]
	\centering
		\includegraphics[width=0.50\textwidth]{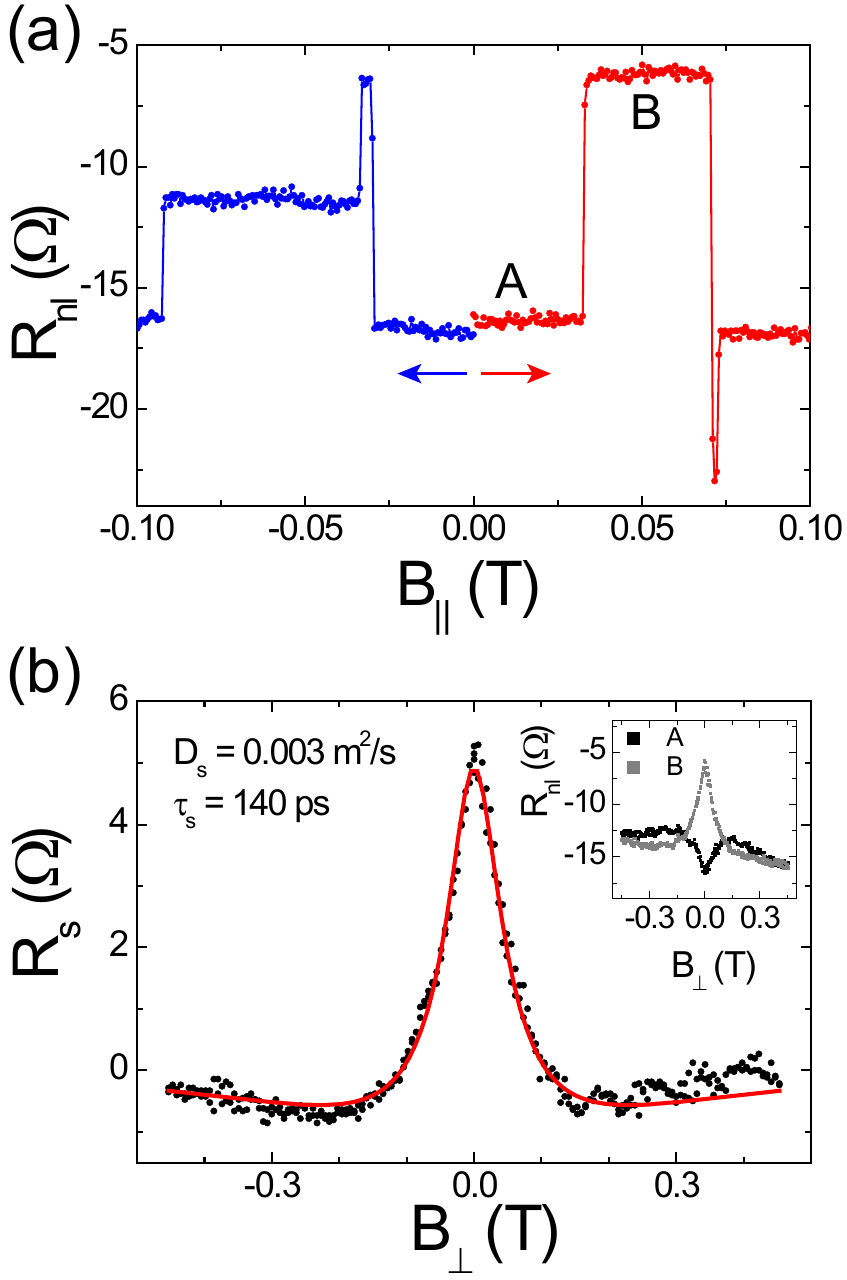}
	\caption{(Color online) (a) Nonlocal spin valve measurement in a graphene quantum dot with $V_{bg}$ = -20 V and $V_{pg}=V_{sg}=$ 0 V at 4.2 K. (b) Nonlocal Hanle precession with the same gate voltages as in (a). The fit using Eq. \ref{eq:oned} is shown by the red line. \textit{Inset}: Data for the nonlocal resistance as a function of a perpendicular magnetic field for the two levels A and B shown in (a).}
	\label{fig:qdspin}
\end{figure}

To check if we can get spin transport through the QD we performed spin valve measurements.
The measurements shown here were performed at 4.2 K.
In Fig. \ref{fig:qdspin}(a) we show that three nonlocal resistance steps, corresponding to the switch of magnetization of three contacts, are visible.
Since contacts 2 and 6 were electrically disconnected from the device, we cannot specify exactly the magnetic configuration of the electrodes for each step.
We have to point out that we only observe a nonlocal spin signal with clear switches when we set $V_{bg}$ to large negative values.
When $V_{bg} >$ 15 V, within the high resistance region of Fig. \ref{fig:qd}, the nonlocal spin signal reduces significantly and the switches get indistinguishable from our measurement noise.

In order to obtain a value for the spin relaxation time in our devices we performed Hanle precession experiments for two different alignment of magnetization of the contacts [levels A and B in Fig. \ref{fig:qdspin}(a)] and obtain the curves $R_{nl}^{A}$ and $R_{nl}^{B}$ [inset of Fig. \ref{fig:qdspin}(b)].
To eliminate background contributions we take the total spin signal as $R_{s} = (R_{nl}^{B}-R_{nl}^{A}) / 2$.
$R_{s}$ is then fitted using Eq. \ref{eq:oned}.
Within the range of $V_{bg}$ = -30 to -20 V we do not see a significant difference in the values obtained, with $\tau_{s} \approx$ 150 ps and $D_{s} \approx$ 0.003 m$^{2}$/s.
As in the case of the spin valve measurements, we could only obtain a Hanle precession signal above the background noise in our device for large negative values of $V_{bg}$.

The value obtained for the spin diffusion coefficient of $D_{s} \approx$ 0.003 m$^{2}$/s indicates the low diffusivity of the QD.
The value for $\tau_{s} \approx$ 150 ps is within the range for the values expected for a standard SiO$_{2}$ based device $\tau_{s} \approx$ 200 ps.
This is no surprise since, as explained in Ref. \cite{WeesNanoLetters2012}, in the case where two outer regions are connected by a central region, the obtained spin relaxation values by Hanle precession can be strongly affected by the outer regions.
This will be elaborated in the section below.

\subsection{Simulations}
\begin{figure}[h]
	\centering
		\includegraphics[width=0.50\textwidth]{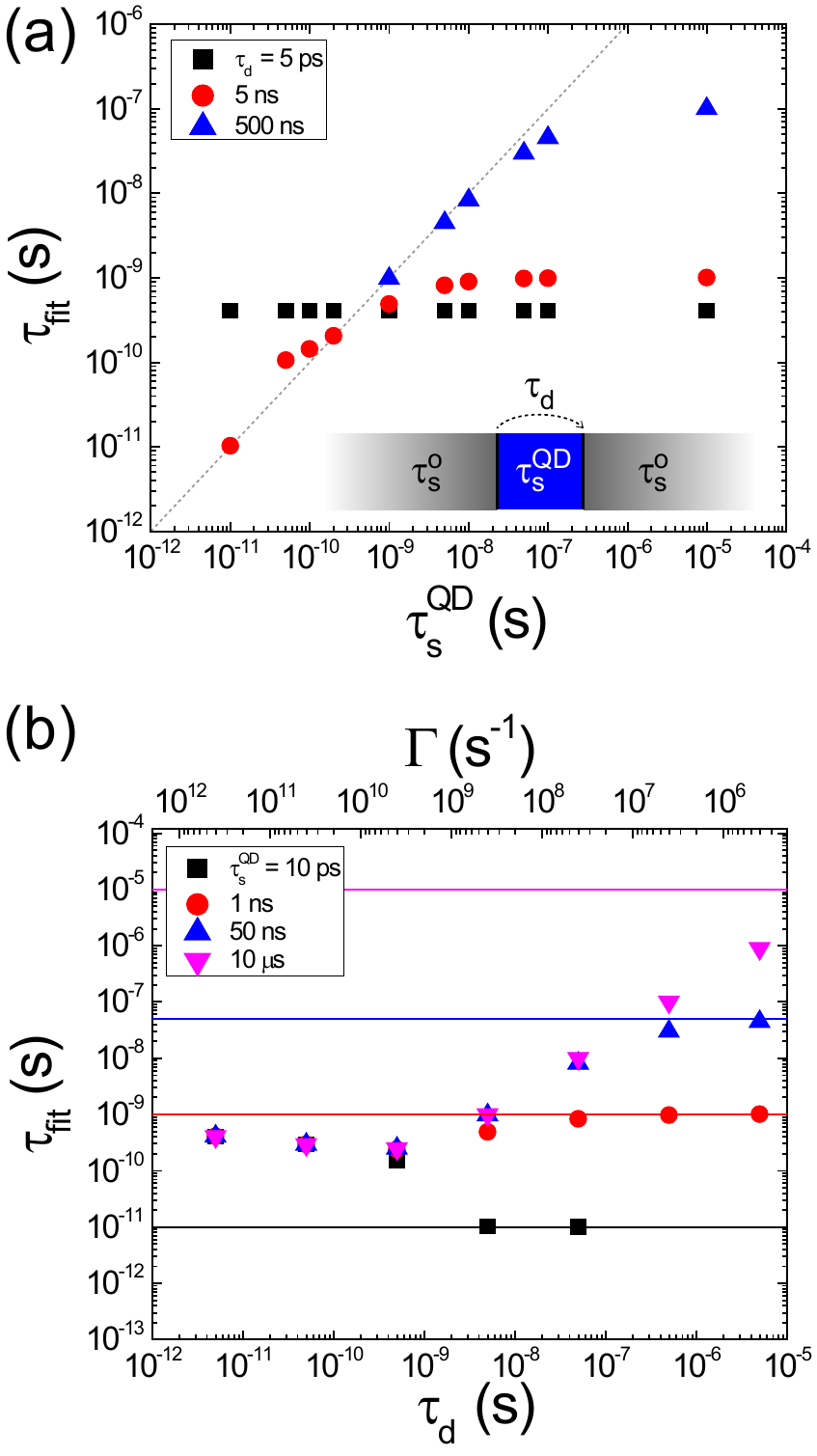}
	\caption{(Color online) (a) $\tau_{fit}$ \textit{versus} $\tau_{s}^{QD}$ for different values of $\tau_{d}$. The dotted line shows $\tau_{fit}=\tau_{s}^{QD}$. \textit{Inset:} Schematics of the system used for the simulations. Two semi-infinite regions with spin relaxation time $\tau_{s}^{o}$ are connected by a central region representing the QD with spin relaxation time $\tau_{s}^{QD}$. The diffusion time between spin injection and detection $\tau_{d}$ is represented by the dotted line. (b) $\tau_{fit}$ \textit{versus} $\tau_{d}$ for different values of $\tau_{s}^{QD}$. The corresponding tunneling rate $\Gamma=2 / \tau_{d}$ is shown in the upper axis. The horizontal lines show the values used for $\tau_{s}^{QD}$.}
	\label{fig:qdsim}
\end{figure}

In order to quantify our results and give a prospect on how to measure the spin relaxation in a QD using Hanle precession measurements, we apply the model developed in Ref. \cite{WeesNanoLetters2012} to our systems.
In this model we map our devices in a system composed of two identical semi-infinite outer regions connected by one inner region of length $\ell$ [see inset of Fig. \ref{fig:qdsim}(a)].
The square resistance, spin relaxation time and spin diffusion coefficient can be set for each region separately.
Here we set the diffusion coefficient and the spin relaxation time for the outer regions as the average values for SiO$_{2}$ based graphene spin valves, $D_{s}^{o} = 0.02$ m$^{2}$/s and $\tau_{s}^{o}$ = 200 ps, respectively.
The spins are injected at the left boundary of the QD and detected at the right boundary.
We then simulate Hanle precession curves by solving Eq. \ref{eq:bloch} with the appropriate boundary conditions and the data is fitted in the same way we do for our experiments, with Eq. \ref{eq:oned}.
From this fitting procedure we extract an effective spin relaxation time for the whole system, $\tau_{fit}$.

The square resistances for the outer regions were set to $R_{o}$ = 1 k$\Omega$ and for the QD $R_{QD}$ = 690 k$\Omega$, and the widths are taken to be the same, $W$ = 100 nm, for simplicity.
Here we use a high value for $R_{QD}$, which would be the case where the QD is in the Coulomb blockade regime.
However, other sets of calculations with different combinations where $R_{o}$ = 1 k$\Omega$ and $R_{QD}$ = 100 k$\Omega$, and $R_{QD} = R_{o}$ show that our conclusions do not depend strongly on the values for $R_{QD}$ and $R_{o}$, if $R_{QD}$ is not several orders of magnitude higher than $R_{o}$.
This fact will be clarified below.

The amount of time the spins spend inside the quantum dot is determined by the tunneling rate ($\Gamma$) of the tunnel barriers between the dot and the graphene leads, created by the narrow graphene ribbons in our experiment.
In our model, we take the two tunnel barriers and the quantum dot as being a single system, and relate the dwell time of the spins in the dot with the spin diffusion time through the barrier/QD/barrier system: $\tau_{d} = 2 / \Gamma$.
The factor of 2 arises from the fact that the spins have to tunnel through two tunnel barriers before reaching the detection circuit.

We start by studying the dependence of $\tau_{fit}$ as a function of the spin relaxation time inside the QD ($\tau_{s}^{QD}$) for different values of $\tau_{d}$.
As depicted in Fig. \ref{fig:qdsim}(a), when $\tau_{d}$ is small, the obtained value for $\tau_{fit}$ is independent on $\tau_{s}^{QD}$, and has a value close to $\tau_{s}^{o}$.
This results from the fact that the spins do not spend sufficient time to experience spin relaxation inside the QD, therefore the obtained value for $\tau_{fit}$ is mostly given by the outside regions.
When the spins spend a longer time inside the dot (larger values of $\tau_{d}$), we observe that $\tau_{fit} \approx \tau_{s}^{QD}$ until $\tau_{s}^{QD} \approx \tau_{d}$.
This can be understood by the fact that, when the spins spend sufficient time inside the dot to relax, the value obtained by fitting the Hanle precession curves represent the spin relaxation inside the QD.

The dependence of $\tau_{fit}$ with $\tau_{d}$ can be understood by realizing that we have a type of conductivity mismatch between the QD and the graphene islands connecting it, similar to the case of a graphene/ferromagnet interface.
We can quantify the mismatch between the QD and the outer regions by the ratio of the spin resistances between both systems.
The spin resistance for the outer regions is given by: $R_{\lambda}^{o} = R_{o} \lambda_{o} / W$, where $\lambda_{o} = \sqrt{D_{o} \tau_{o}}$ is the spin relaxation length in the outer regions.
Using $\lambda_{s}^{QD} = \sqrt{\tau_{s}^{QD} (2 \ell^{2} / \tau_{d})}$, the spin resistance for the dot can be written as a function of the length between the two tunnel barriers $\ell$, the diffusion time and the spin relaxation in the dot as: $R_{\lambda}^{QD} = (\ell R_{QD} / W \sqrt{2}) \times \sqrt{\tau_{s}^{QD} / \tau_{d}}$.
Therefore, the ratio between the spin resistances is: $R_{\lambda}^{o} / R_{\lambda}^{QD} = (R_{o} / R_{QD}) \times (\lambda_{o} / \ell) \times \sqrt{2 \tau_{d} / \tau_{s}^{QD}}$.
In the case of small values of $\tau_{d}$ and/or large values of $\tau_{s}^{QD}$, the ratio $R_{\lambda}^{o} / R_{\lambda}^{QD}$ is small and there is a high mismatch between the regions.
Therefore the obtained spin relaxation time, $\tau_{fit}$, is given by the outer regions.
For the case of long spin diffusion times, the mismatch is smaller and the values obtained for the spin relaxation time are more representative of the relaxation in the QD.
This is a type of impedance mismatch in which an important parameter is the spins' time of the flight through the QD compared to their spin relaxation inside the dot.

In order to understand what would be necessary to detect the spin relaxation in graphene quantum dots with the theoretically predicted $\tau_{s}^{QD} \approx$ 10 $\mu$s \cite{BurkardNatPhys2007}, we study the behavior of $\tau_{fit}$ with $\tau_{d}$.
Our analysis is summarized in Fig. \ref{fig:qdsim}(b).
It can be seen that the spin relaxation in the QD starts to have an influence on the obtained $\tau_{fit}$ when $\tau_{d} \approx \tau_{s}^{o}$.
For low values of $\tau_{s}^{QD}$, soon after $\tau_{d} \approx \tau_{s}^{o}$, $\tau_{fit}$ decreases sharply and saturates at $\tau_{s}^{QD}$.
For long spin relaxation times in the QD when compared to the outside regions, $\tau_{fit}$ increases slowly with the increase of $\tau_{d}$ and saturates at $\tau_{fit} = \tau_{s}^{QD}$ for $\tau_{d} > 10 \tau_{s}^{QD}$.

Translating the diffusion time of the spins through our structure to a tunneling rate in/out the QD, it can be seen that, in order to measure the spin relaxation times predicted for graphene QDs using Hanle precession, we require highly decoupled QDs, with $\Gamma < 10^{5}$ s$^{-1}$.
Although these values for the tunneling rate are very low, values of $\Gamma \approx$ 10$^{5}$ s$^{-1}$ were experimentally demonstrated in graphene QDs \cite{IhnPhys.Rev.B2011}.
The value of $\Gamma$ can be easily tuned in graphene QDs by a local $V_{sg}$ \cite{EnsslinMaterialsToday2010}.
Therefore, we expect that by studying the values of spin relaxation time obtained using Hanle precession measurements as a function of $\Gamma$ would reveal the intrinsic spin relaxation time in graphene QDs.

It is important to point out however, that when the diffusion time is very long the spin signal is very low.
For our simulations with $\tau_{d}>$ 5 ns the total amplitude of the simulated spin signal was too small to be fitted.
This happens due to two effects.
First, the spins have time to relax in the dot, which reduces the total signal.
And second, as explained above, the spin resistance in the direction across the dot is much higher than in the direction away from the dot.
Consequently, the spins tend to diffuse away in the opposite direction and very few spins travel across the dot and are detected.
Therefore, for the type of studies presented here, it is necessary to increase the nonlocal spin signal by, for example, increasing the spin polarization of the contacts \cite{ValenzuelaAppliedPhysicsLetters2013,KawakamiPhys.Rev.Lett.2010} and at the same time increase the time the spins spend inside the QD by decreasing the tunneling rate.

In our experiments $\tau_{d} = \ell ^{2} / 2 D_{s} \approx$ 75 ps, with $\ell$ = 0.63 $\mu$m, and the obtained spin relaxation time is $\approx$ 150 ps.
When we compare this value with Fig. \ref{fig:qdsim}(b), we see that the value for the spin relaxation extracted using Hanle precession is invariant with respect to $\tau_{s}^{QD}$.
This means that our measurements are dominated by the spin relaxation in the outer regions.
As stated above, in order to obtain a value closer to the value for the spin relaxation time in the QD we have to combine contacts with high spin polarization with a QD weakly coupled to the graphene islands (outer regions).

\section{Conclusions}
In conclusion we showed spin accumulation and transport in graphene nanostructures.
We demonstrated the effect of confinement in graphene nanoislands with dimensions smaller than the spin relaxation length.
By Hanle precession measurements we could extract the spin relaxation time in these systems.
Using a theoretical model, we showed that for these 0D systems the effect of contact induced spin relaxation is much higher than for the standard devices where spins can diffuse away.
When the contact resistance is sufficiently high to not induce spin relaxation, the maximum value for the nonlocal spin signals in confined systems is more than one order of magnitude higher than for unbound systems, where the total length is much longer than the spin relaxation length.
Comparing simulations and experiments we see that the low experimentally obtained values for $\tau_{s}$ seem to be due to contact induced spin relaxation.
Furthermore, by using the experimental values in our simulations, we see that $\tau_{s}$ in our graphene nanoislands has to be considerably higher than 20 ps in order to match the values obtained by our experiments.
Therefore the spin relaxation in the graphene nanoislands do not seem to be reduced by the higher edge length to area ratio in our samples, which indicates a low influence of enhanced spin-flip mechanisms at the graphene edges.

We also studied the nonlocal spin transport in a graphene quantum dot connected by two graphene nanoislands.
In this case, the contact resistances were high enough in order to reduce significantly the effect of contact induced spin relaxation.
A value for $\tau_{s} \approx$ 150 ps was obtained by Hanle precession measurements.
By simulating our devices we showed that this value for the spin relaxation time seems to be due to spin relaxation in the outer graphene islands and not by the quantum dot due to the short time the spins spend inside the quantum dot.
We explain this effect using by estimating the spin resistance mismatch between the outer graphene islands and the graphene QD.
Our simulations indicate that, in order to obtain a more representative value for $\tau_{s}$ in quantum dots using nonlocal Hanle precession measurements, one should increase the time the spins spend inside the quantum dot, which can be achieved by reducing the tunneling rate through the tunnel barriers that connect the QD.
However, the nonlocal spin signal reduces significantly since the spins tend to diffuse away from the QD and very few make it through and are detected on the other side of the dot.
Therefore, the use contacts with high spin polarization in combination with a highly decoupled quantum dot should allow for the extraction of the spin relaxation time inside the QD using nonlocal Hanle precession measurements.

\section{Acknowledgments}
We would like to acknowledge J. G. Holstein, H. M. de Roosz, B. Wolfs and H. Adema for the technical support.

The research leading to these results has received funding from the Dutch Foundation for Fundamental Research on Matter (FOM), the European Union Seventh Framework Programme under grant agreement n$^{\circ}$604391 Graphene Flagship, the Netherlands Organisation for Scientific Research (NWO), the People Programme (Marie Curie Actions) of the European Union's Seventh Framework Programme FP7/2007-2013/ under REA grant agreement n$^{\circ}$607904-13 Spinograph, NanoNed and the Zernike Institute for Advanced Materials.

\bibliography{nanostructures-paper}

\end{document}